\documentclass[fina,3p, onecolumn]{elsarticle}
\usepackage{color}
\usepackage{hyperref}
\usepackage[svgnames]{xcolor}

\journal{Journal of \LaTeX\ Templates}
\usepackage{natbib}
\newcommand{\Th}{\Theta}
\newcommand{\om}{\omega}
\newcommand{\Om}{\Omega}

\newcommand{\ep}{\epsilon}
\newcommand{\si}{\sigma}

\newcommand{\de}{\delta}

\newcommand{\ra}[1]{\renewcommand{\arraystretch}{#1}}

\newcommand{\nn}{\nonumber}

\newcommand{\fr}{\frac}

\newcommand{\updown}{\uparrow\downarrow}

\newcommand{\bt}{\textbf}

\usepackage{amsmath, bm, mathtools,amssymb,makeidx,array, booktabs}
\usepackage{geometry,graphicx,subcaption,wasysym}
\usepackage[hang,flushmargin]{footmisc}
\usepackage[justification=centering]{caption}
\usepackage{array}
\newcolumntype{L}[1]{>{\raggedright\let\newline\\\arraybackslash\hspace{0pt}}m{#1}}
\newcolumntype{C}[1]{>{\centering\let\newline\\\arraybackslash\hspace{0pt}}m{#1}}
\newcolumntype{R}[1]{>{\raggedleft\let\newline\\\arraybackslash\hspace{0pt}}m{#1}}

\DeclarePairedDelimiter\abs{\lvert}{\rvert}%
\makeatletter
\let\oldabs\abs
\def\abs{\@ifstar{\oldabs}{\oldabs*}}

\newcommand{\Eq}[1]{Eq.~(\ref{#1})}

\newcommand{\vr}{\bm{r}}

\begin{document}

\begin{frontmatter}

\title{Polarized EMC Effect in the QMC Model\tnoteref{prep}}

\tnotetext[prep]{CSSM preprint ADP-18-14/T1062}

\author[cssm]{S.~Tronchin}
\author[cssm]{H.~H.~Matevosyan\corref{cor1}}
\author[cssm]{A.~W.~Thomas}

\cortext[cor1]{Corresponding author}

\address[cssm]{CSSM and ARC Centre of Excellence for Particle Physics at the Terascale, Department of Physics, University of Adelaide SA 5005 Australia}

\begin{abstract}
The ratios of the in-medium to free nucleon structure functions for the unpolarized and polarized cases are obtained using the MIT bag model for the free case, along with the QMC model to incorporate the in-medium modifications of the structure functions. As discussed in earlier work, the observed nuclear EMC effect is reasonably well described. This gives us confidence to investigate the predictions of the model for the polarized EMC effect, the ratio $g_1^*(x)/g_1(x)$ for a bound proton to that of a free proton. This ratio is found to be substantially different from unity and very similar in shape and size  to that found in the unpolarized case. This prediction of such a fundamental change in the valence structure of a bound nucleon needs to be tested experimentally at the earliest opportunity.
\end{abstract}


\end{frontmatter}


\section{Introduction}
Our understanding of nuclear structure was severely challenged by the unexpected experimental results released by the European Muon Collaboration (EMC) in 1983. Roughly speaking, EMC compared the structure function of a free nucleon to that of a bound nucleon and found there was a significant difference between them~\cite{Aubert:1983xm,Arnold:1983mw}, especially in the valence region. The ratio of the in-medium to free structure function of the nucleon was found to drop below one in that region, indicating a suppression of the bound nucleon's structure function, which became known as the EMC effect~\cite{Geesaman:1995yd,Malace:2014uea,Norton:2003cb}. 

The discovery that the nucleon structure functions differ substantially in-medium compared to the free case surprised the nuclear physics community, pointing to a potential new approach to nucleon structure. One such approach is the Quark Meson Coupling (QMC) 
model~\cite{Guichon:1987,Fleck:1990td,Guichon:1995ue}, which explicitly allows the quark degrees of freedom to respond self-consistently to the nuclear mean fields and leads naturally to changes in the internal structure of the bound nucleons. Initial work within the QMC model did in fact reproduce the main features of the EMC effect~\cite{Thomas:1989vt,Saito:1992rm} but this work required some finesse to overcome technical difficulties with momentum conservation in the bag model~\cite{Schreiber:1991tc}. 

More recently Clo\text{\"e}t, Bentz and Thomas~\cite{Cloet:2005rt,Cloet:2006bq} overcame the limitations of the bag model by using the covariant NJL model~\cite{Nambu:1961tp} for the structure of the nucleon within a QMC-like framework in which, once again, the internal structure of the nucleon was self-consistently solved in the mean fields generated in nuclear matter~\cite{Bentz:2001vc}. The calculations made in this model were successful in reproducing the unpolarized EMC data across the periodic table. These authors also extended their calculations to make predictions for  the spin dependent  structure function of odd-A nuclei. There they found a considerably larger change, referred to as the ''polarized EMC effect'', than that found in the unpolarized case.

Here we aim to investigate the model dependence of the results reported by 
Clo\text{\"e}t {\it et al.} by calculating the spin dependent EMC effect in the QMC model, using the MIT bag model~\cite{Bogolubov:1968zk,Chodos:1974je,Chodos:1974pn,Thomas:1982kv} for the structure of the nucleon. This is especially important as there is yet no consensus on the origin of the EMC effect. The work done within the QMC model and its NJL generalization is based upon mean-field theory, with the structure of all of the bound nucleons modified in proportion to the local mean fields. On the other hand, there has recently been much interest in the suggestion that the EMC effect might only be associated with those nucleons involved in short-range 
correlations~\cite{Hen:2016kwk,Hen:2013oha,Weinstein:2010rt}. As the contribution of correlated nucleons to the spin dependent structure function of a nucleus is expected to be suppressed, a measurement of this effect promises to be a valuable source of information in separating these two explanations. In parallel with such a measurement, we need the very best estimates of the polarized EMC effect within any particular model. 

\section{Calculation of the Structure Functions}

In this work we employ the MIT bag model~\cite{Bogolubov:1968zk,Chodos:1974je,Chodos:1974pn,Thomas:1982kv} to describe the structure of the free nucleon. It is a simple, yet successful phenomenological model for quark confinement, within which  three non-interacting quarks are bound to a spherical region of space, with the boundary condition that the quark vector current normal to the surface vanishes. The quarks being treated as non-interacting is justified by appealing to the idea of asymptotic freedom, and the hard boundary condition is a crude implementation of quark confinement. An attractive advantage of the model is that many calculations can be carried through analytically, giving valuable insight into the origin of the effects of the medium on the quark distributions.

As we already remarked, while the usual formulation of the bag model is convenient for calculating static properties, it is not ideal for the study of deep inelastic scattering where the conservation of energy and momentum is essential to ensuring the correct support of the parton distribution functions (PDFs). For this reason Schreiber {\em et al.}~\cite{Schreiber1991} chose to perform the calculations of the PDFs in a framework where energy and momentum conservation was 
imposed~\cite{Jaffe:1985je} before any approximations were made. The formulation of Schreiber {\em et al.} builds in these conservation laws and, in addition, uses the Peierls-Yoccoz 
method~\cite{Peierls:1957er} to construct approximate momentum eigenstates. 
 
 The PDF of a longitudinally polarized quark of flavor $f$ inside of a longitudinally polarized nucleon of mass $M$ can be calculated by evaluating:
\begin{align} \label{qf_ini}
  q_f^{\updown}(x) = \sum_n  \delta\left(1 -x - \frac{p_n^+}{P^+} \right) \left |  \langle n | \Psi_{+,f} | P,s \rangle \right|^2 ,
\end{align}
where $x$ is the light-cone ($+$)\footnote{The $+$ component of a 4-vector $a$ here is defined as $a^+ = a^0 + a^3$} momentum fraction of the nucleon carried by the quark, which is described by wave function $\Psi_{f}$, while $P$ and $s$ describe the momentum and the spin of the nucleon. The  notation  $\updown$ indicates aligned or anti-aligned helicities of the quark and the nucleon and the polarized PDF can be expressed (e.g.,~\cite{Barone:2001sp}) in terms of the conventional unpolarized, $f(x)$, and the helicity dependent, $\Delta f(x)$, distributions as
%
$q^{\updown}(x) = {1}/{2} ( f(x) \pm \Delta f(x)  )$.
The sum is over all possible intermediate states $n$ with momenta $p_n$ when probing the quark in a deep inelastic scattering process. We consider only  diquark intermediate states, that provide the dominant contribution to the quark PDF~\cite{Schreiber1991}, which yields the following expression
\begin{align} \label{qf}
q_f^{\updown}(x) = \fr{M}{(2\pi)^2}\sum_m \langle \mu|P_{f,m}|\mu \rangle \int_{\fr{\abs{M^2(1-x)^2-M_n^2}}{2M(1-x)}}^\infty p_ndp_n\fr{|\phi_2(\bt{p}_n)|^2}
{|\phi_3(\bt{0})|^2} |\tilde{\Psi}_m^{\updown}(\bt{p}_n)|^2 \, ,
\end{align}
The spin-flavor wave function of the initial nucleon at rest is denoted by $|\mu \rangle$, and $P_{f,m}$ is a projector operator onto quark of flavor $f$ and spin $m$. The functions $\phi_3$ and $\phi_2$ are the normalizations of the 3-quark and diquark wave functions, and $M_n$ is the mass of the intermediate state.

In the zero mass quark case the MIT bag wave function for a spatial coordinate $\bm{r}$ takes the form \cite{Chodos:1974je,Thomas1982kv}
\begin{equation} \label{Psi}
\Psi_m(\vr) = N 
\begin{pmatrix}
j_0 \left( \fr{\Om |\vr|}{R} \right) \chi_m \\
i \bm{\sigma} \cdot \bm{\hat{r}} \ j_1 \left( \fr{\Om |\vr|}{R} \right) \chi_m
\end{pmatrix} 
\Th (R-|\vr|),
\end{equation}
with lowest energy eigenfrequency solution $\Om \simeq 2.04$. Here $R$ is the bag radius, $j_0$ and $j_1$ are the spherical Bessel functions of the first kind, $\chi_m$ are spinors, and $\bm{\sigma}$ are the Pauli spin matrices. The normalization of the wave function is given by 
\begin{align} \label{N2}
N^2 = \fr{1}{4\pi} \fr{\Om^3}{2R^3(\Om-1)\sin^2(\Om)}.
\end{align}

Using the Peierls-Yoccoz method to obtain the approximate eigenstates of momentum, we find
\begin{align} \label{phi2 eq}
|\phi_2(\bt{p}_n)|^2 = \fr{4\pi R}{u} \left( \fr{2\pi N^2 R^4}{\Om^4} \right)^2 \int \fr{dv}{v} \sin\left(\fr{2vu}{\Om} \right)T^2(v)
\end{align}
and
\begin{align} \label{phi3 eq}
|\phi_3(\bt{0})|^2 = 4\pi \left( \fr{2\pi N^2 R^4}{\Om^4} \right)^3 \int \fr{dv}{v} T^3(v) \, ,
\end{align}
where $T(v)$ is the overlap function for quarks in displaced bags, and the following substitutions have also been made
\begin{align} \label{vu sub}
v = \fr{|\vr|\Om}{2R} \;, \quad u = |\bt{p}_n|R\;.
\end{align}

The overlap function can be evaluated using the bag wave function, yielding
\begin{align} \label{T}
T(v) &= T_t(v) + T_b(v),
 \\ 
T_t(v)&= \left[ (\Om-v)\sin(2v)+(1-\sin^2(\Om))-\cos(\Om)\cos(\Om-2v) \right] ,
\\ 
T_b(v)&=\Bigg[ \left(1-\fr{4v^2}{2\Om^2} \right)\sin^2(\Om) - \fr{2}{\Om}\sin(\Om)\cos(\Om-2v) + \fr{2}{\Om}\sin(\Om)\cos(\Om)+ \Om \sin(2v)
 \\ \nn
&\quad \quad - \sin(\Om)\sin(\Om-2v) - v\sin(2v) \Bigg],
\end{align}
where $T_t(v)$ corresponds to the overlap integral of the upper component of the bag wave function and $T_b(v)$ to the lower component. 

The Fourier transform of $\Psi$ is given as 
\begin{align} \label{FT}
|\tilde{\Psi}_m^{\updown}(\bt{p}_n)|^2 = \fr{1}{2} \left[f(\bt{p}_n) \pm (-1)^{m+3/2} g(\bt{p}_n) \right],
\end{align}
where 
\begin{align} \label{fpn}
f(\bt{p}_n) = \fr{\pi R^3}{2} \fr{\Om^3}{\left(\Om^2-\sin^2(\Om) \right)} \left[ s_1^2(u)+2\fr{p_n^z}{|\bt{p}_n|} s_1(u)s_2(u)+s_2^2(u) \right],
\end{align}
and
\begin{align} \label{gpn}
g(\bt{p}_n) = \fr{\pi R^3}{2} \fr{\Om^3}{\left(\Om^2-\sin^2(\Om) \right)} \left\{ s_1^2(u)+2\fr{p_n^z}{|\bt{p}_n|} s_1(u)s_2(u)+ \left[1-2\left(\fr{p_n^{\perp}}{|\bt{p}_n|}\right)^2\right]s_2^2(u) \right\}.
\end{align}
Here we have defined
\begin{align} \label{pnz}
p_n^z = M(1-x) - \sqrt{\bt{p}_n^{2}+M_n^2} \, ,
\end{align}
\begin{align} \label{pnp}
p_n^{\perp 2} = 2M(1-x) \sqrt{M_n^2+\bt{p}_n^2} - M^2(1-x)^2-M_n^2 \, ,
\end{align}
and the function
\begin{align} \label{s1}
s_1(u) = \fr{1}{u} \left[ \fr{\sin(u-\Om)}{u-\Om} - \fr{\sin(u+\Om)}{u+\Om} \right],
\end{align}
corresponds to the upper component of the bag wave function, while the function
\begin{align} \label{s2}
s_2(u) = 2j_0(\Om)j_1(u) - \fr{u}{\Om}s_1(u),
\end{align}
corresponds to the lower component of the bag wave function.

The polarized PDF in \Eq{qf}  for $u$ and $d$ quarks can be then evaluated using the spin-flavor matrix elements taken from Ref.~\cite{Schreiber1991}
\begin{align} \label{udpol}
u^{\updown}(x) &= F(x) \pm \frac{2}{3} G(x),
\\
d^{\updown}(x) &= \frac{1}{2}F(x) \mp \frac{1}{6} G(x),
\end{align}
where
\begin{align} \label{F2}
F_{}(x) &= \fr{M}{(2\pi)^2} \int_{\fr{\abs{M^2(1-x)^2-M_n^2}}{2M(1-x)}}^\infty p_ndp_n\fr{|\phi_2(\bt{p}_n)|^2}{|\phi_3(\bt{0})|^2} \ f(\bt{p}_n) ,
\end{align}
and
\begin{align} \label{G2}
G_{}(x) &= \fr{M}{(2\pi)^2} \int_{\fr{\abs{M^2(1-x)^2-M_n^2}}{2M(1-x)}}^\infty p_ndp_n\fr{|\phi_2(\bt{p}_n)|^2}{|\phi_3(\bt{0})|^2} \ g(\bt{p}_n)
\end{align}
 are obtained by substituting the spin-dependence decomposition of $\Psi$ in \Eq{FT} into the expression~(\ref{qf}).

Phenomenologically, it is important to include the effect of the hyperfine splitting arising from one-gluon-exchange in the calculations of the quark distributions. This accounts for the mass splitting between a diquark state with quark spins aligned, the so-called vector intermediate state with the higher mass ($M_{n,v}$), or anti-aligned, the so-called scalar intermediate state with the lower mass ($M_{n,s}$). The fact that the vector intermediate state is about 50 MeV heavier than a diquark without hyperfine splitting and the scalar intermediate state is about $150~\mathrm{MeV}$ lighter, explains many of the spin and flavor dependent features of the observed PDFs~\cite{Close:1988}. The quark distributions are then given by
\begin{align} \label{u gluon}
u_{}^{\updown}(x) &= \left[ \fr{3}{4}F_{s}(x)+\fr{1}{4}F_{v}(x) \right] \pm \fr{2}{3} \left[ \fr{9}{8}G_{s}(x)-\fr{1}{8}G_{v}(x) \right],
\end{align}
and
\begin{align} \label{d gluon}
d_{}^{\updown}(x) &= \fr{1}{2}F_{v}(x) \mp \fr{1}{6}G_{v}(x).
\end{align}
The subscripts $s$ and $v$ indicate that the diquark masses $M_{n,s}$ and $M_{n,v}$, respectively, are to be used when evaluating the distributions $F_{s}(x)$ and $F_{v}(x)$  (Eq.~(\ref{F2})) and $G_{s}(x)$ and $G_{v}(x)$ (Eq.~(\ref{G2})). \\

The contributions from the four-quark intermediate states can be determined in a similar way to the two-quark intermediate state~\cite{Schreiber1991}. However, the additional uncertainties associated with the four-quark contributions, which are confined to small-$x$, lead us to replace an explicit calculation of the shape by the phenomenological form $(1-x)^7$, which is very close to that calculated for $F_{(4)}(x)$ in Ref.~\cite{Schreiber1991}. 

\subsection{In-medium effects}

For simplicity, we consider an isoscalar medium in which we need only the scalar and  the vector mean fields. The latter simply shifts energy scales, while the former modifies the mass of the confined quark leading to significant changes in the valence quark wave function. In the nuclear medium, the mean scalar field,  $\bar{\sigma}$, modifies the quark mass as $m_q^* = m_q - g_\sigma^q \bar{\sigma}$, where $m_q$ is the mass of the quark inside a free nucleon, and $g_\sigma^q$ is the coupling of the scalar field to the light quarks. The quark wave function in the bound nucleon at the spatial coordinate $\bm{r}$ is given by~\cite{Thomas1982kv}
\begin{equation}
\Psi_m^*(\vr) = N^{*2} 
\begin{pmatrix}
j_0 \left( \fr{\Om |\vr|}{R} \right) \chi_m \\
i b \bm{\sigma} \cdot \bm{\hat{\vr}} \ j_1 \left( \fr{\Om |\vr|}{R} \right) \chi_m
\end{pmatrix} 
\Th (R-|\vr|),
\end{equation}
where 
\begin{align}
b &= \left( \fr{E-m_q^*}{E+m_q^*} \right)^{\fr{1}{2}}, \\
E &= \fr{1}{R} \left( \Om^{2}+(m_q^* R)^2 \right)^{\fr{1}{2}},
\end{align}
and $N^*$ is the new normalization factor.
The factor $b$, appearing in the lower component of the bag wave function, leads to an alteration of the expression for $s_2(u)$ (c.f. Eq.~(\ref{s2})), so that
\begin{align} \label{s2*}
s_2^*(u) = b \left( 2j_0(\Om)j_1(u) - \fr{u}{\Om}s_1(u) \right) \, ,
\end{align}
while the expression for $s_1(u)$ (Eq. (\ref{s1})) remains unchanged. The functions arising from the Peierls-Yoccoz projection become
\begin{align} \label{phi2 si}
|\phi_2(\bt{p}_n)|^{*2} = \fr{4\pi R}{u} \left( \fr{2\pi N^{*2} R^4}{\Om^4} \right)^2 \int \fr{dv}{v} \sin\left(\fr{2vu}{\Om}\right) T^{*2}(v),
\end{align}
and
\begin{align} \label{phi3 si}
|\phi_3(\bt{0})|^{*2} = 4\pi \left( \fr{2\pi N^{*2} R^4}{\Om^4} \right)^3 \int \fr{dv}{v} T^{*3}(v) \, ,
\end{align}
where the overlap integral of the bag wave function is now 
\begin{align}
T^*(v) &= T_t + b^2T_b,
\end{align}
where $T_t$ and $T_b$ have the same functional form as in the free case (Eq.~\ref{T}). In order to determine the eigenfrequency one must solve the equation
\begin{align}
\tan(\Om) = \fr{ \Om }{ 1-m_q^* R-\left( \Om^2+(m_q^* R)^2 \right)^{\fr{1}{2}} }.
\end{align}

Apart from the change in the valence quark wave function, the effect of the $\si$ mean field also needs to be included through effective masses of the nucleon and the diquarks. The effective mass of a bound nucleon is given by~\cite{Guichon:2018uew}
\begin{align} \label{Meff}
M^*(\bar{\si}) = M - g_{\si}\bar{\si} + \fr{d}{2} \left(g_{\si}\bar{\si}\right)^2,
\end{align}
where $d=0.22R$, and $R$ is the bag radius. Studies have shown that Eq. (\ref{Meff}) is quite accurate up to values of $g_{\si}\bar{\si}=400$ MeV~\cite{Guichon:2004xg}, which will be sufficient for our purposes. \\

Including the full effect of the $\si$ mean field the quark distribution is now given by
\begin{align} \label{qf si}
q_{N_0}^{\updown}(x) = \fr{M^*}{(2\pi)^2}\sum_m \langle \mu|P_{f,m}|\mu \rangle \int_{{\fr{\abs{M^{*2}(1-x)^2-M_n^{*2}}}{2M^*(1-x)}}}^\infty p_ndp_n\fr{|\phi_2(\bt{p}_n)|^{*2}}{|\phi_3(\bt{0})|^{*2}} |\tilde{\Psi}_m^{*\updown}(\bt{p}_n)|^2,
\end{align} 
where the subscript 0 indicates that only the scalar field has been included and the vector field is yet to be incorporated. \\

\subsection{Fermi Motion}
The effect of Fermi motion on a bound nucleon can be included through a convolution of the quark distribution with a Fermi smearing function, $f_0(\tilde{y}_A)$, which is given by~\cite{Mineo:2003vc}
\begin{align} \label{fermi}
f_0(\tilde{y}_A) = \fr{3}{4} \left(\fr{E_F}{p_F}\right)^3 \left[ \left(\fr{p_F}{E_F}\right)^2 - (1-\tilde{y}_A)^2 \right],
\end{align}
where the distribution has support for
\begin{align}
1-\fr{p_F}{E_F} < \tilde{y}_A < 1 + \fr{p_F}{E_F}.
\end{align}
The Fermi energy is given by
\begin{align}
E_F = \sqrt{p_F^2+M^{*2}},
\end{align}
where $p_F$ is the Fermi momentum. Performing the convolution of the quark distribution in Eq. (\ref{qf si}) with $f_0(\tilde{y}_A)$ yields the distribution
\begin{align}
q_{A_0}(\tilde{x}_A) = \int d\tilde{y}_A \int dx \; \de(\tilde{x}_A-\tilde{y}_Ax) \; q_{N_0}(x) f_0(\tilde{y}_A).
\end{align}
Eliminating the $\de$-function we obtain the result
\begin{align} \label{qA0}
q_{A_0}(\tilde{x}_A) = \int d\tilde{y}_A \; \fr{1}{\tilde{y}_A} \; q_{N_0}\left(\fr{\tilde{x}_A}{\tilde{y}_A}\right) f_0(\tilde{y}_A) \, ,
\end{align}
where, again, the subscript 0 indicates that the vector field effects have not yet been included. \\

\subsection{The $\om$ Mean Field}
The vector field is included by scaling the quark distribution $q_{A_0}(\tilde{x}_A)$, and shifting the Bjorken variable $x$. The in-medium quark distribution then becomes~\cite{Mineo:2003vc}
\begin{align} \label{qA}
q_A(x_A) = \fr{\ep_F}{E_F} q_{A_0} \left( \tilde{x}_A = \fr{\ep_F}{E_F}x_A - \fr{V_0}{E_F} \right),
\end{align}
where
\begin{align}
\ep_F = \sqrt{p_F^2+M^{*2}} + 3V_0 \equiv E_F+3V_0
\end{align}
is the Fermi energy with the vector field included. Here $V_0$ is the mean vector potential felt by each light quark (hence $3V_0$ for the nucleon as a whole). The new variable $x_A$ is the Bjorken scaling variable for a nucleon in the nucleus. These distributions may be plotted as a function of $x$ through the relation $x_A = \fr{M}{\ep_F}x$~\cite{Cloet:2005tq}. 

\section{Results}
\begin{figure}[h!]
	\centering
	\captionsetup{justification=raggedright}
	\includegraphics[scale=0.4]{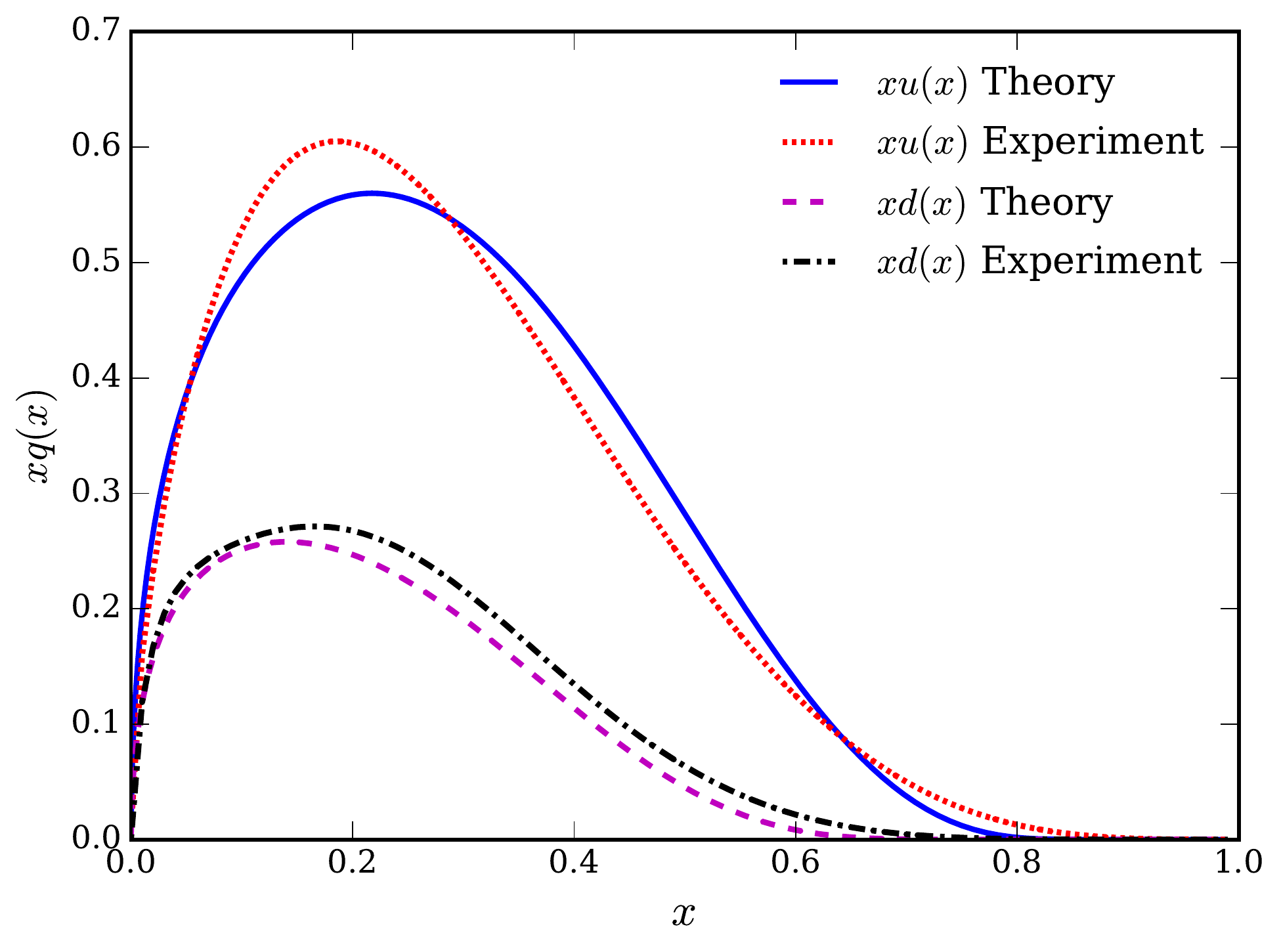}
	\caption{A comparison of the up and down nucleon valence PDFs calculated within the bag model, after NLO evolution to $Q^2=10$ GeV$^2$, with the NLO distributions of 
Martin {\it et al.}~\cite{Martin:2009iq}. The red dotted and black dash-dotted lines depict the up and down valence results Martin {\it et al.} (Experiment), while the blue solid and purple dashed lines depict the corresponding results in our model (Theory).}
	\label{fig:uandd}
\end{figure}

 As usual, in order to compare the  PDFs calculated in a valence quark model to those measured in experiment, we must identify a non-perturbative scale, $\mu$, at which the model best approximates the physical nucleon. This scale may be thought of as part of the model. It is determined by fitting one observable, for example the unpolarized structure function $F_2^p(x)$ at $5~\mathrm{GeV}^2$~\cite{Schreiber:1990ij,Diakonov:1998ze,Diakonov:1996sr}. Here, the model scale is set $\mu = 0.2~\mathrm{GeV^2}$ by fitting the free, unpolarized valence quark PDFs after evolution with the next-to-leading-order QCD evolution equations, by utilizing QCDNUM software tool~\cite{Botje:2010ay}. It is worth noting, that the results presented here do not change significantly when using the leading order QCD evolution equations, and following the same procedure for determining the (different) model scale. Figure~\ref{fig:uandd} shows a  comparison between the model and the phenomenological nucleon valence PDFs of Martin {\it et al.}~\cite{Martin:2009iq} at 10 GeV$^2$. Clearly the model reproduces the main features of the empirical distributions quite well, although they do tend to go to zero a little too fast at 
large $x$.

\begin{figure}[h!]
	\centering
	\captionsetup{justification=raggedright}
	\includegraphics[scale=0.4]{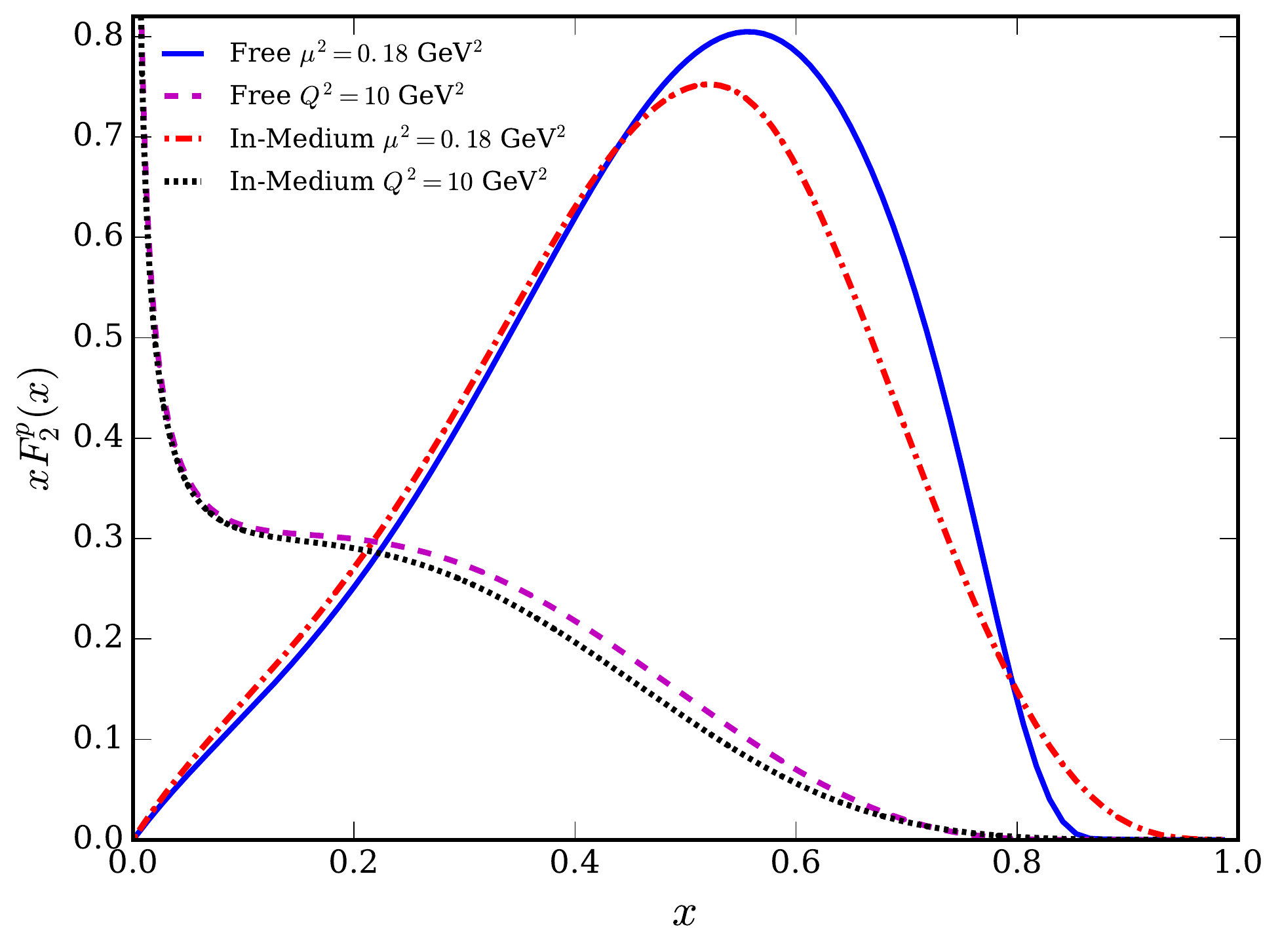}
	\caption{Spin independent structure functions for the proton in the bag model. The solid blue line describes the free nucleon results at the model scale, while the purple dashed line depicts the results evolved to $Q^2=10~\mathrm{Gev}^2$. The red dash-dotted and the black dotted liens are the corresponding results for the in-medium nucleon.}
	\label{fig:xF2p_med_free}
\end{figure}

\begin{figure}[h!]
	\centering
	\captionsetup{justification=raggedright}
	\includegraphics[scale=0.4]{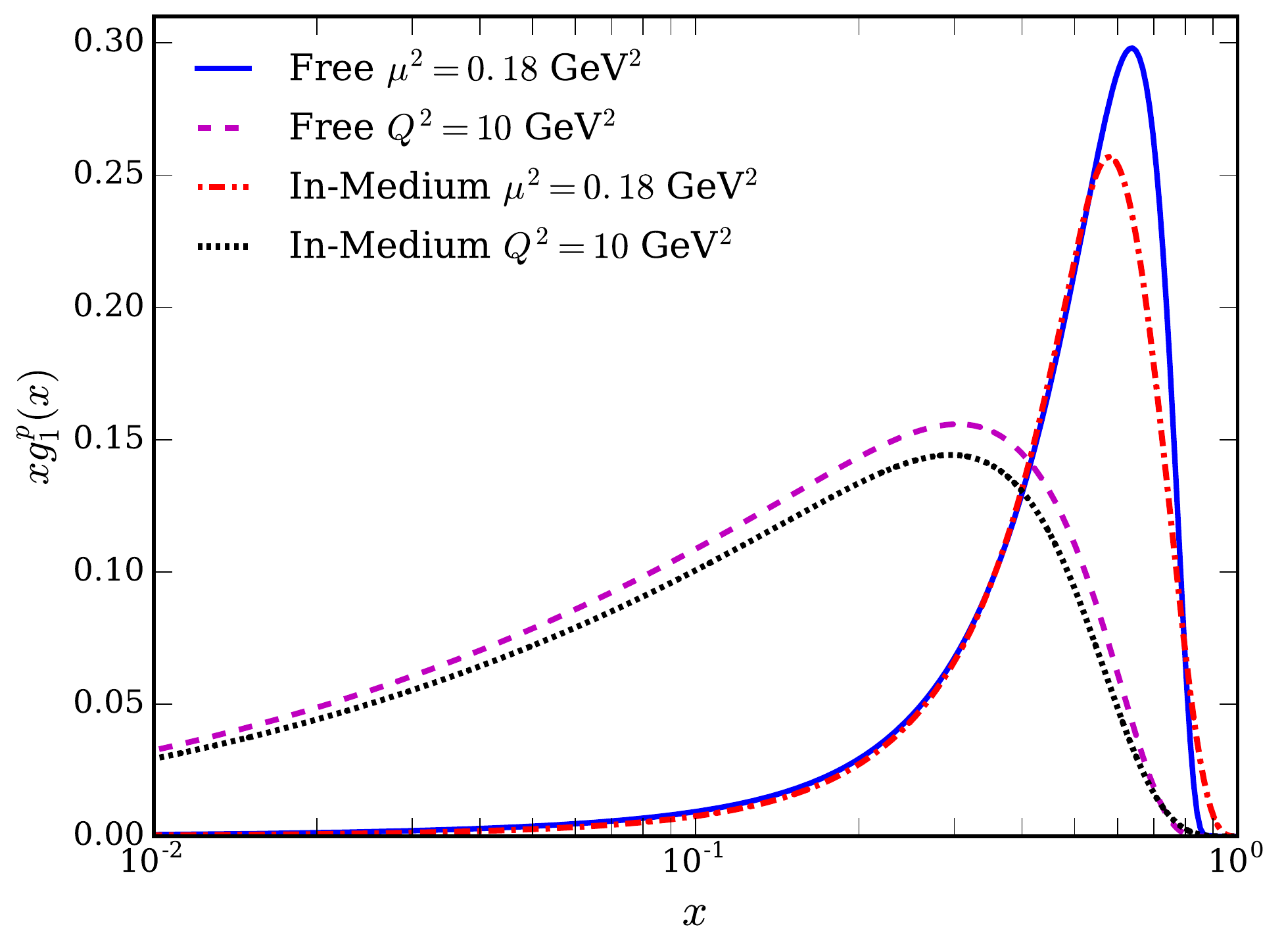}
	\caption{Spin dependent structure functions for the proton in the bag model.  The solid blue line describes the free nucleon results at the model scale, while the purple dashed line depicts the results evolved to $Q^2=10~\mathrm{Gev}^2$. The red dash-dotted and the black dotted liens are the corresponding results for the in-medium nucleon.}
	\label{fig:xg1p_med_free}
\end{figure}

\begin{figure}[h!]
	\centering
	\captionsetup{justification=raggedright}
	\includegraphics[scale=0.4]{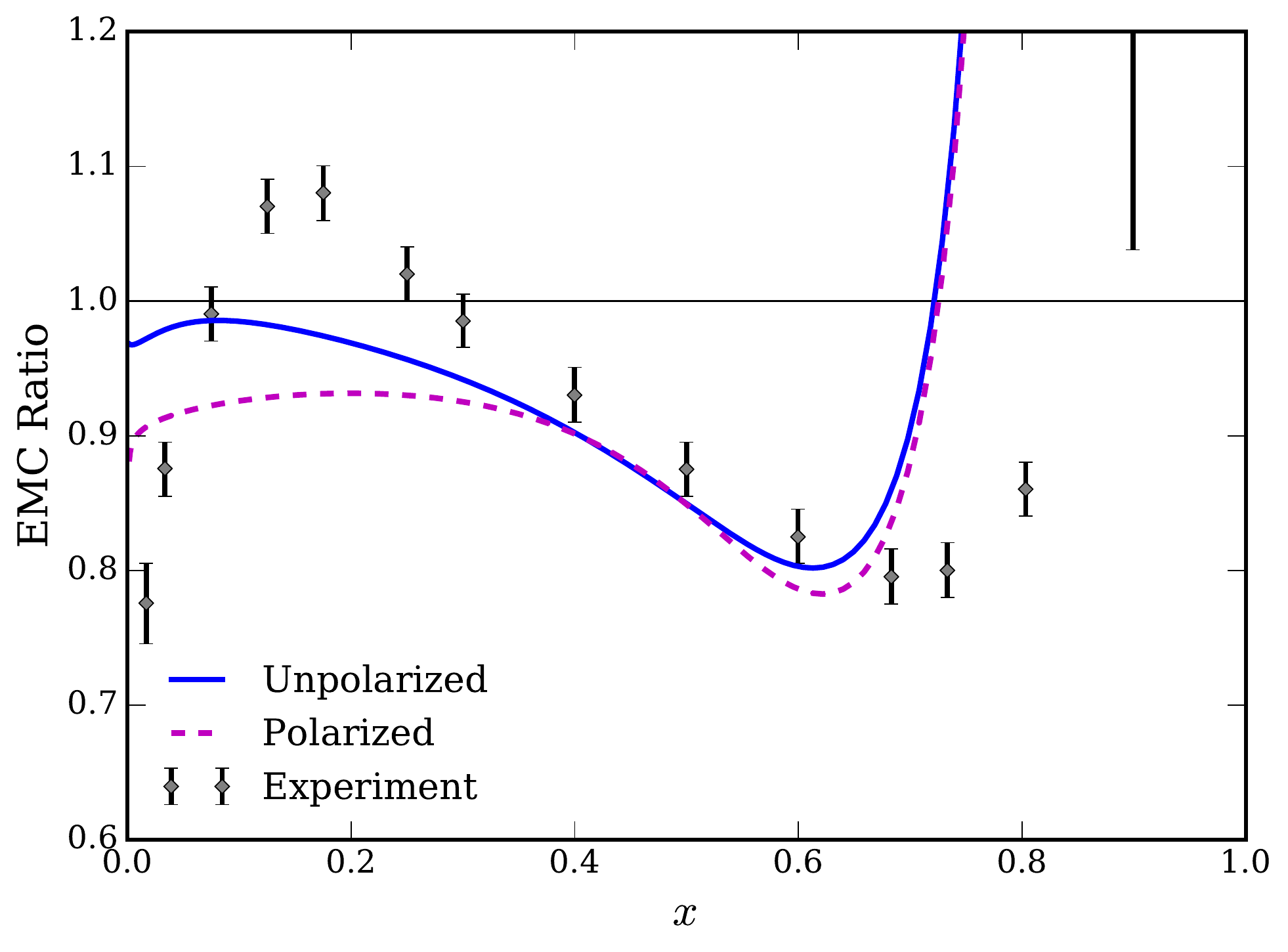}
	\caption{Unpolarized (blue solid line) and polarized (purple dashed line) EMC effect in the QMC model. The results are evolved to $Q^2=10~\mathrm{GeV}^2$. The unpolarized EMC experimental data for nuclear matter is taken from Ref.~\cite{Sick:1992pw}.}
	\label{fig:EMC_Com}
\end{figure}

The parameters used here are summarized in Table~\ref{tab:bestfit}. The values taken for 
the $\sigma$ and $\omega$ mean fields are typical of those found in the literature~\cite{Guichon:1995ue,Saito:2005rv}.We present the in-medium and free structure functions of the proton for the unpolarized and polarized case in Fig.~\ref{fig:xF2p_med_free} and Fig.~\ref{fig:xg1p_med_free}, respectively. We then present both the unpolarized and polarized EMC ratios for nuclear matter together in Fig.~\ref{fig:EMC_Com}. 

\begin{table}[h!]
	\centering
	\ra{1.2}
	\begin{tabular}{ ccccc }
		\toprule
		\multicolumn{5}{c}{Set Values} \\
		\midrule
		$R$ (fm) & $g_{\si}\bar{\si}$ (MeV)  & $M$ & $M_{n,s}$ & $M_{n,v}$ \\
		1.0 & 255  & 938.3 & 500 & 700 \\ 
		\bottomrule
	\end{tabular}
	\\ \vspace{0.2cm}
	\begin{tabular}{ cccccc }
		\toprule
		\multicolumn{6}{c}{Resultant Values} \\
		\midrule
		$m_q^*$ & $\Om$ & $V_0$ (MeV) & $M^*$ &  $M_{n,s}^*$ & $M_{n,v}^*$ \\
		-175 & 1.42 & 62 & 720 & 354 & 554 \\
		\bottomrule
	\end{tabular}
	\captionsetup{justification=raggedright}
	\caption{Parameter set used for the calculations of the PDFs. All the masses are in units of MeV.}
	\label{tab:bestfit}
\end{table}

In the process of obtaining our results we were able to determine which in-medium modifications play the dominant role in producing the EMC effect. We found that the altered quark wave function, which arises both from the $\si$ mean field, as well as the effect of Fermi motion, plays a very minor role in producing the EMC effect. The key players in producing the EMC effect, for both the unpolarized and polarized case, are the effect of the $\si$ mean field on the nucleon and diquark masses and the energy shifts associated with the $\om$ mean field. The effective nucleon and diquark state masses, arising from the $\si$ mean field, cause a suppression of the in-medium structure function in the low to mid-$x$ region. The vector potential, arising from the $\om$ mean field, suppresses the in-medium structure function in the mid to high-$x$ region, and also slightly opposes the effect of the $\si$ mean field in the low-$x$ region. Together these effects play the dominant role in producing the EMC effect. The premature rise in the EMC ratio at large $x$ is related to the deficiency of the bag model for which the PDFs go to zero too rapidly in this region, as noted earlier.

From the result presented in Fig.~\ref{fig:EMC_Com}, we see that the prediction of the polarized EMC effect in the QMC model is about the same as that of the unpolarized effect.  We stress that this polarized EMC effect is defined for a proton which is 100\% polarized. In a real nucleus, such as $^7$Li, for which a measurement is planned at Jefferson Lab~\cite{JLAB:PR12-14-001}, one needs to account for the fact that the polarization of the bound proton is less than that of the nucleus. 

Clearly there is a substantial EMC correction predicted for the spin dependent structure function, as large as the spin independent EMC effect. On the other hand, it is somewhat smaller than the spin dependent effect predicted in the NJL model~\cite{Cloet:2005rt}. This is almost certainly related to the much stronger suppression of $g_A$ found by Clo\"et {\it et al.}, which was of order 20\% in nuclear matter, roughly twice as large as the 8.9\% suppression found in the present work, with similar values in earlier QMC calculations~\cite{Guichon:1987,Saito:2005rv}. In view of the importance of any suppression of $g_A$ for searches for double 
beta-decay~\cite{Vergados:2016hso}, where it enters to the fourth power, this deserves further study.

\section{Conclusion}
We have studied the unpolarized and polarized EMC effect using the MIT bag to model the free nucleon and the mean field approximation along with the QMC model to model the bound nucleon. The in-medium modifications included the effect of the $\si$ (scalar) and $\om$ (vector) mean fields and Fermi motion.

We found that the calculation of the unpolarized EMC effect describes the trend of the experimental data in the valence region. The calculation also yielded a sizable polarized EMC effect, with the prediction being very similar in magnitude and shape to that of the unpolarized effect. We also found that together, the effective nucleon and diquark state masses 
arising from the $\si$ mean field, and the vector potential arising from the $\om$ mean field are the dominant contributors in producing the EMC effect for both the unpolarized and polarized case. 

Our results on the relative sizes of the unpolarized and polarized EMC effects significantly differ with those previous calculations in the NJL model~\cite{Cloet:2005rt}, while it seems likely that the polarized effect is altogether suppressed in the short-range correlations approach~\cite{Hen:2016kwk,Hen:2013oha}. The upcoming experiments at Thomas Jefferson National Laboratory are expected to measure the polarized EMC effect in the near future, which should provide valuable new evidence concerning the different mechanisms for generating both the unpolarized and polarized EMC effects. This would provide us with further insight into the structure of the nucleon, as there is no doubt that spin physics has historically played a crucial role in developing our understanding of hadron structure.

\section*{Acknowledgements}
This work was supported by the University of Adelaide and by the Australian Research Council through the ARC Centre of Excellence for Particle Physics at the Terascale (CE110001104) and Discovery Project DP150103164.


\bibliographystyle{ieeetr}
\bibliography{paper}

\begin{thebibliography}{10}

\bibitem{Aubert:1983xm}
J.~J. Aubert {\em et~al.}, ``{The ratio of the nucleon structure functions
  $F2_n$ for iron and deuterium},'' {\em Phys. Lett.}, vol.~B123, pp.~275--278,
  1983.

\bibitem{Arnold:1983mw}
R.~G. Arnold {\em et~al.}, ``{Measurements of the a-Dependence of Deep
  Inelastic electron Scattering from Nuclei},'' {\em Phys. Rev. Lett.},
  vol.~52, p.~727, 1984.

\bibitem{Geesaman:1995yd}
D.~F. Geesaman, K.~Saito, and A.~W. Thomas, ``The nuclear emc effect,'' {\em
  Ann. Rev. Nucl. Part. Sci.}, vol.~45, pp.~337--390, 1995.

\bibitem{Malace:2014uea}
S.~Malace, D.~Gaskell, D.~W. Higinbotham, and I.~Cloet, ``{The Challenge of the
  EMC Effect: existing data and future directions},'' {\em Int. J. Mod. Phys.},
  vol.~E23, no.~08, p.~1430013, 2014.

\bibitem{Norton:2003cb}
P.~R. Norton, ``{The EMC effect},'' {\em Rept. Prog. Phys.}, vol.~66,
  pp.~1253--1297, 2003.

\bibitem{Guichon:1987}
P.~A.~M. Guichon, ``{A Possible Quark Mechanism for the Saturation of Nuclear
  Matter},'' {\em Phys. Lett.}, vol.~B200, pp.~235--240, 1988.

\bibitem{Fleck:1990td}
S.~Fleck, W.~Bentz, K.~Yazaki, and K.~Shimizu, ``{A Sigma omega quark model to
  saturate nuclear matter},'' {\em Nucl. Phys.}, vol.~A510, pp.~731--739, 1990.

\bibitem{Guichon:1995ue}
P.~A.~M. Guichon, K.~Saito, E.~N. Rodionov, and A.~W. Thomas, ``{The Role of
  nucleon structure in finite nuclei},'' {\em Nucl. Phys.}, vol.~A601,
  pp.~349--379, 1996.

\bibitem{Thomas:1989vt}
A.~W. Thomas, A.~Michels, A.~W. Schreiber, and P.~A.~M. Guichon, ``{A NEW
  APPROACH TO NUCLEAR STRUCTURE FUNCTIONS},'' {\em Phys. Lett.}, vol.~B233,
  pp.~43--47, 1989.

\bibitem{Saito:1992rm}
K.~Saito, A.~Michels, and A.~W. Thomas, ``{Towards a microscopic understanding
  of nuclear structure functions},'' {\em Phys. Rev.}, vol.~C46,
  pp.~R2149--R2152, 1992.

\bibitem{Schreiber:1991tc}
A.~W. Schreiber, A.~I. Signal, and A.~W. Thomas, ``{Structure functions in the
  bag model},'' {\em Phys. Rev.}, vol.~D44, pp.~2653--2662, 1991.

\bibitem{Cloet:2005rt}
I.~C. Cloet, W.~Bentz, and A.~W. Thomas, ``{Spin-dependent structure functions
  in nuclear matter and the polarized EMC effect},'' {\em Phys. Rev. Lett.},
  vol.~95, p.~052302, 2005.

\bibitem{Cloet:2006bq}
I.~C. Cloet, W.~Bentz, and A.~W. Thomas, ``{EMC and polarized EMC effects in
  nuclei},'' {\em Phys. Lett.}, vol.~B642, pp.~210--217, 2006.

\bibitem{Nambu:1961tp}
Y.~Nambu and G.~Jona-Lasinio, ``{Dynamical Model of Elementary Particles Based
  on an Analogy with Superconductivity. 1.},'' {\em Phys. Rev.}, vol.~122,
  pp.~345--358, 1961.
\newblock [,127(1961)].

\bibitem{Bentz:2001vc}
W.~Bentz and A.~W. Thomas, ``{The Stability of nuclear matter in the
  Nambu-Jona-Lasinio model},'' {\em Nucl. Phys.}, vol.~A696, pp.~138--172,
  2001.

\bibitem{Bogolubov:1968zk}
P.~N. Bogolubov, ``{On a Model of quasiindependent quarks},'' {\em Ann. Inst.
  H. Poincare Phys. Theor.}, vol.~8, pp.~163--190, 1968.

\bibitem{Chodos:1974je}
A.~Chodos, R.~L. Jaffe, K.~Johnson, C.~B. Thorn, and V.~F. Weisskopf, ``{A New
  Extended Model of Hadrons},'' {\em Phys. Rev.}, vol.~D9, pp.~3471--3495,
  1974.

\bibitem{Chodos:1974pn}
A.~Chodos, R.~L. Jaffe, K.~Johnson, and C.~B. Thorn, ``{Baryon Structure in the
  Bag Theory},'' {\em Phys. Rev.}, vol.~D10, p.~2599, 1974.

\bibitem{Thomas:1982kv}
A.~W. Thomas, ``{Chiral Symmetry and the Bag Model: A New Starting Point for
  Nuclear Physics},'' {\em Adv. Nucl. Phys.}, vol.~13, pp.~1--137, 1984.

\bibitem{Hen:2016kwk}
O.~Hen, G.~A. Miller, E.~Piasetzky, and L.~B. Weinstein, ``{Nucleon-Nucleon
  Correlations, Short-lived Excitations, and the Quarks Within},'' {\em Rev.
  Mod. Phys.}, vol.~89, no.~4, p.~045002, 2017.

\bibitem{Hen:2013oha}
O.~Hen, D.~W. Higinbotham, G.~A. Miller, E.~Piasetzky, and L.~B. Weinstein,
  ``{The EMC Effect and High Momentum Nucleons in Nuclei},'' {\em Int. J. Mod.
  Phys.}, vol.~E22, p.~1330017, 2013.

\bibitem{Weinstein:2010rt}
L.~B. Weinstein, E.~Piasetzky, D.~W. Higinbotham, J.~Gomez, O.~Hen, and
  R.~Shneor, ``{Short Range Correlations and the EMC Effect},'' {\em Phys. Rev.
  Lett.}, vol.~106, p.~052301, 2011.

\bibitem{Schreiber1991}
A.~W. Schreiber, A.~I. Signal, and A.~W. Thomas, ``Structure functions in the
  bag model,'' {\em Phys. Rev. D}, vol.~44, pp.~2653--2662, Nov 1991.

\bibitem{Jaffe:1985je}
R.~L. Jaffe, ``{Deep inelastic scattering with application to nuclear
  targets},'' in {\em {Proceedings, Research Program at CEBAF I: Report of the
  1985 Summer Study Group, June 10 - August 30, 1985}}, 1985.

\bibitem{Peierls:1957er}
R.~E. Peierls and J.~Yoccoz, ``{The Collective model of nuclear motion},'' {\em
  Proc. Phys. Soc.}, vol.~A70, pp.~381--387, 1957.

\bibitem{Barone:2001sp}
V.~Barone, A.~Drago, and P.~G. Ratcliffe, ``{Transverse polarisation of quarks
  in hadrons},'' {\em Phys. Rept.}, vol.~359, pp.~1--168, 2002.

\bibitem{Thomas1982kv}
A.~W. Thomas, ``{Chiral Symmetry and the Bag Model: A New Starting Point for
  Nuclear Physics},'' {\em Adv. Nucl. Phys.}, vol.~13, pp.~1--137, 1984.

\bibitem{Close:1988}
F.~E. Close and A.~W. Thomas, ``{The Spin and Flavor Dependence of Parton
  Distribution Functions},'' {\em Phys. Lett.}, vol.~B212, pp.~227--230, 1988.

\bibitem{Guichon:2018uew}
P.~A.~M. Guichon, J.~R. Stone, and A.~W. Thomas, ``{Quark–Meson-Coupling
  (QMC) model for finite nuclei, nuclear matter and beyond},'' {\em Prog. Part.
  Nucl. Phys.}, vol.~100, pp.~262--297, 2018.

\bibitem{Guichon:2004xg}
P.~A.~M. Guichon and A.~W. Thomas, ``{Quark structure and nuclear effective
  forces},'' {\em Phys. Rev. Lett.}, vol.~93, p.~132502, 2004.

\bibitem{Mineo:2003vc}
H.~Mineo, W.~Bentz, N.~Ishii, A.~W. Thomas, and K.~Yazaki, ``{Quark
  distributions in nuclear matter and the EMC effect},'' {\em Nucl. Phys.},
  vol.~A735, pp.~482--514, 2004.

\bibitem{Cloet:2005tq}
I.~C. Cloet, W.~Bentz, and A.~W. Thomas, ``{Spin-dependent parton distributions
  in the nucleon},'' {\em Nucl. Phys. Proc. Suppl.}, vol.~141, pp.~225--232,
  2005.
\newblock [,225(2005)].

\bibitem{Schreiber:1990ij}
A.~W. Schreiber, A.~W. Thomas, and J.~T. Londergan, ``{{QCD} Evolution of the
  Spin Structure Functions of the Neutron and Proton},'' {\em Phys. Rev.},
  vol.~D42, pp.~2226--2236, 1990.

\bibitem{Diakonov:1998ze}
D.~Diakonov, V.~{\relax Yu}. Petrov, P.~V. Pobylitsa, M.~V. Polyakov, and
  C.~Weiss, ``{On Nucleon parton distributions from the chiral quark soliton
  model},'' {\em Phys. Rev.}, vol.~D58, p.~038502, 1998.

\bibitem{Diakonov:1996sr}
D.~Diakonov, V.~Petrov, P.~Pobylitsa, M.~V. Polyakov, and C.~Weiss, ``{Nucleon
  parton distributions at low normalization point in the large N(c) limit},''
  {\em Nucl. Phys.}, vol.~B480, pp.~341--380, 1996.

\bibitem{Botje:2010ay}
M.~Botje, ``{QCDNUM: Fast QCD Evolution and Convolution},'' {\em Comput. Phys.
  Commun.}, vol.~182, pp.~490--532, 2011.
\newblock
  \url{https://www.nikhef.nl/user/h24/qcdnum-files/doc/qcdnum170008.pdf}.

\bibitem{Martin:2009iq}
A.~D. Martin, W.~J. Stirling, R.~S. Thorne, and G.~Watt, ``{Parton
  distributions for the LHC},'' {\em Eur. Phys. J.}, vol.~C63, pp.~189--285,
  2009.

\bibitem{Saito:2005rv}
K.~Saito, K.~Tsushima, and A.~W. Thomas, ``{Nucleon and hadron structure
  changes in the nuclear medium and impact on observables},'' {\em Prog. Part.
  Nucl. Phys.}, vol.~58, pp.~1--167, 2007.

\bibitem{Sick:1992pw}
I.~Sick and D.~Day, ``{The EMC effect of nuclear matter},'' {\em Phys. Lett.},
  vol.~B274, pp.~16--20, 1992.

\bibitem{JLAB:PR12-14-001}
W.~Brooks {\em et~al.}, Jefferson Lab experiment proposal PR12-14-001.
\newblock \url{https://www.jlab.org/exp_prog/proposals/14/PR12-14-001.pdf}.

\bibitem{Vergados:2016hso}
J.~D. Vergados, H.~Ejiri, and F.~Šimkovic, ``{Neutrinoless double beta decay
  and neutrino mass},'' {\em Int. J. Mod. Phys.}, vol.~E25, no.~11, p.~1630007,
  2016.

\end{thebibliography}

\end{document}